\documentclass[conference]{IEEEtran}
\IEEEoverridecommandlockouts

\usepackage{cite}
\usepackage{amsmath,amssymb,amsfonts}
\usepackage{graphicx}
\usepackage[caption=false,font=footnotesize]{subfig}
\usepackage{textcomp}
\usepackage{xcolor}
\usepackage{url}
\usepackage{balance}


\def\BibTeX{{\rm B\kern-.05em{\sc i\kern-.025em b}\kern-.08em
    T\kern-.1667em\lower.7ex\hbox{E}\kern-.125emX}}

\begin{document}

\title{An LLM-Orchestrated Agent Framework for CKM-Assisted Network Deployment Optimization}

\author{
	\IEEEauthorblockN{Yilong Chen\textsuperscript{1,2}, Yuan Guo\textsuperscript{1,2}, and Jie Xu\textsuperscript{1,2}}
	\IEEEauthorblockA{
		\textsuperscript{1}School of Science and Engineering (SSE), The Chinese University of Hong Kong (Shenzhen), Shenzhen, China\\
		\textsuperscript{2}Shenzhen Future Network of Intelligence Institute (FNii-Shenzhen), Shenzhen, China\\
		Email: yilongchen@link.cuhk.edu.cn, guoyuan@cuhk.edu.cn, xujie@cuhk.edu.cn}
	\thanks{The work was supported in part by the Guangdong Major Project of Basic and Applied Basic Research (No. 2023B0303000001), the Basic Research Project No. HZQB-KCZYZ-2021067 of Hetao Shenzhen-HK S\&T Cooperation Zone, the National Natural Science Foundation of China under grants Nos. U25A20390 and 62471424, the Guangdong Provincial Key Laboratory of Future Networks of Intelligence under grant No. 2022B1212010001, and the Shenzhen Fundamental Research Program under grant No. JCYJ20250604141209012.}
	\thanks{Jie Xu is the corresponding author.}
}

\maketitle

\begin{abstract}
	Channel knowledge maps (CKMs) offer reusable, site-specific radio priors to facilitate environment-aware network planning. 
	However, deploying heterogeneous network nodes to meet diverse performance requirements in complex environments is challenging and requires executable workflows that coordinate scene preparation, candidate-site specification, CKM construction, optimization, and validation.
	This paper presents a large language model (LLM)-orchestrated agent framework for CKM-assisted network deployment optimization, where an LLM translates high-level requirements into modular, typed, and reproducible planning workflows. 
	In this framework, an LLM layer interprets intent, selects modules, and populates normalized configurations; physics-grounded modules generate scene, CKM, optimization, and validation artifacts; and a deterministic guarded controller performs validation-guided refinement.
	A common target-candidate-CKM-optimizer interface supports the registration and composition of compatible planning modules, with CKM backends, samplers, and solvers remaining replaceable. 
	We instantiate the proposed framework for base station (BS) deployment, intelligent reflecting surface (IRS) deployment, and unmanned aerial vehicle (UAV) trajectory planning. An urban-block case study demonstrates automatic scene consistency checks, bounded BS/IRS deployment refinement, and collision- and radio-feasible UAV planning.
\end{abstract}


\section{Introduction}

Sixth-generation (6G) wireless networks are expected to deploy dense infrastructure and diverse network nodes, such as on-ground and aerial base stations (BSs), intelligent reflecting surfaces (IRSs), and unmanned aerial vehicles (UAVs), to support ubiquitous connectivity, enhanced artificial intelligence (AI) capabilities, aerial users, integrated sensing and communication (ISAC), reconfigurable propagation environments, and intelligent services in complex site-specific environments \cite{tong20216g}. This growing diversity of nodes, services, and environments makes network planning increasingly complex. Practical deployment may require jointly placing BSs, selecting IRS mounting walls, and identifying low-altitude UAV corridors while meeting coverage, reliability, cost, and geometric constraints. Conventional approaches based on statistical channel models provide only coarse coverage footprints, while manual configuration and task-specific scripts make them difficult to adapt and scale across heterogeneous sites and objectives.

To provide the site-specific propagation information absent from such statistical models, channel knowledge maps (CKMs) have emerged as a promising abstraction for environment-aware network planning \cite{zeng2021ckm,ckm_background}. A CKM query may return channel power gain, yielding a channel gain map (CGM), or location-dependent descriptors such as line-of-sight (LoS) states, delay-angle features, or beam information. Depending on the planning task, a base-station-to-any (B2X) CKM fixes the BS position and is indexed by receiver position, whereas an any-to-any (X2X) CKM allows both link endpoints to vary \cite{biwgs}. CKMs can be constructed through interpolation of sparse measurements, environment-informed image-processing or generative AI methods, and wireless radiance field (WRF) approaches \cite{ckm_construction_recent}. Beyond channel mapping, SRCON uses white-box and machine-learning models to reproduce traffic arrivals, user association, and signal propagation in real-world cellular networks, enabling data-driven performance evaluation and optimization \cite{srcon}. Site-specific channel knowledge has also been exploited for IRS placement \cite{irs_deployment} and joint low-altitude air-corridor and BS planning \cite{uav_planning}. These studies demonstrate the value of environment- and data-aware models, but focus on prescribed models or individual applications, leaving the orchestration of heterogeneous planning tasks largely unaddressed.

\begin{figure*}[!tb]
	\centering
	\includegraphics[width=0.75\textwidth]{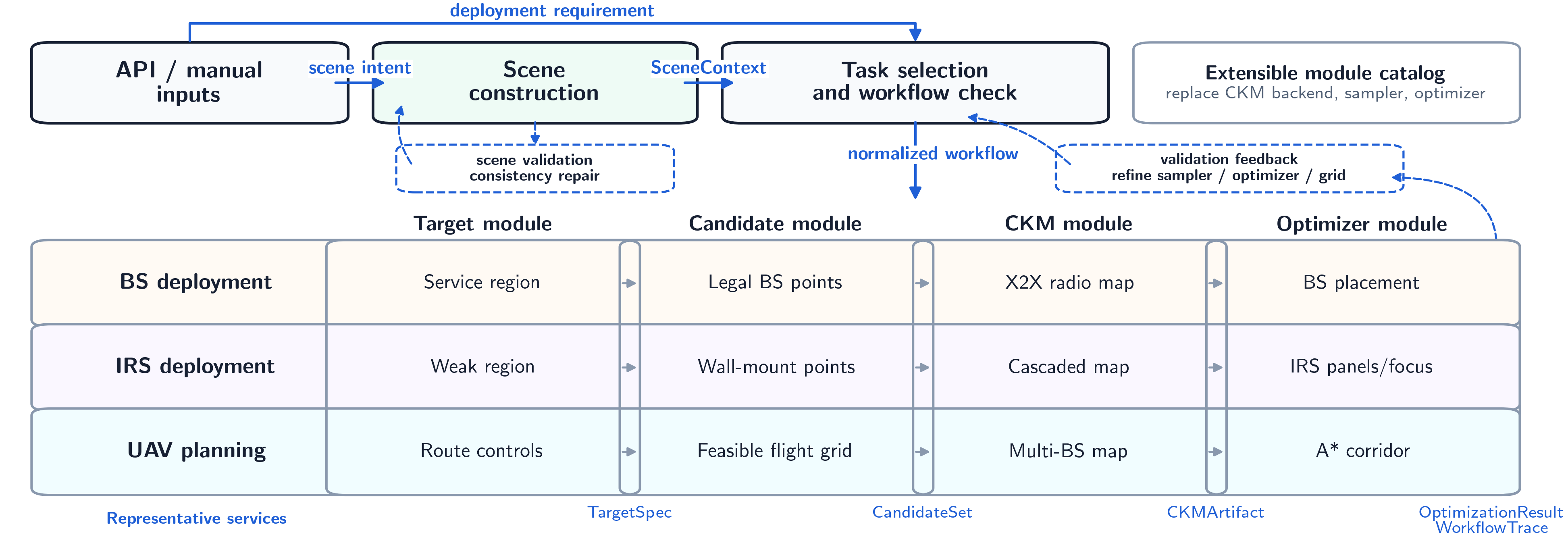}
	\caption{Operation flow and typed module interfaces of the LLM-orchestrated agent framework. The LLM composes an initial workflow from the deployment requirement; deterministic scene validation and guarded deployment refinement wrap the target, candidate, CKM, and optimizer stages connected by typed artifacts. The three rows show representative service instantiations of the extensible interface.}
	\label{fig:workflow}
\vspace{-0.2cm}\end{figure*}

Despite this progress, conventional CKM-assisted planning systems often use rigid, task-specific pipelines, requiring manual configuration and software integration whenever the scene, propagation model, optimization objective, or deployment service changes. By contrast, large language model (LLM) agents offer a promising alternative because, beyond interpreting high-level requirements, they can invoke external software tools, coordinate multistep operations, and use intermediate observations to improve task performance \cite{llm_wireless_survey,wirelessagent}. For example, the LLM agent in \cite{llm_agents_radio_map} calls scene-construction, CKM-generation, and BS-optimization tools to automate a labor-intensive planning process. The framework in \cite{maintained} further uses schema-defined function calls to coordinate geographic data processing, propagation analysis, and mathematical optimization from a single natural-language objective. These works establish intent-driven, tool-grounded network planning, but remain tied to the implemented services, with limited support for composing or extending the planning logic. An extensible workflow paradigm is therefore needed to make planning modules interoperable and allow user-defined deployment services to be integrated without redesigning the agent.

To fill this gap, we develop an LLM-orchestrated agent framework that turns high-level natural-language deployment requirements into modular, typed, and reproducible workflows for CKM-assisted planning. The main contributions are threefold. First, we propose an orchestration architecture in which the LLM layer interprets deployment intent, selects services and modules, and populates their configurations, while physics-grounded modules perform scene processing, CKM construction, and numerical optimization. This separation enables natural-language interaction while keeping physical modeling and numerical computation outside the LLM. Second, we define typed contracts across scene preparation, target specification, candidate generation, CKM construction, optimization, and validation. The contracts allow CKM backends, samplers, and solvers to be registered, substituted, and composed, while deterministic guards check scene consistency, perform bounded validation-guided refinement, and preserve a reproducible workflow trace. Third, we implement BS placement, IRS deployment, and UAV trajectory planning on the Sionna RT backend \cite{sionna}; an urban-block case study demonstrates automatic scene checking, bounded deployment refinement, and radio- and collision-feasible trajectory planning. The source code is available at \url{https://github.com/chenyilong-cuhksz/ckm-planning-agent}.

\section{Agent Workflow and Typed Interfaces}

Fig.~\ref{fig:workflow} shows the operation flow of the proposed LLM-orchestrated agent framework. A user configures access to an LLM application programming interface (API) and provides a scene model file. Scene-preparation prompts guide the LLM-assisted interface in turning the file into a radio scene, which deterministic checks validate. The user then selects a registered planning service and states a deployment requirement. From this requirement, the LLM composes an initial workflow across four stages, namely target definition, candidate generation, CKM construction, and optimization, and populates the corresponding module configurations. Typed artifacts feed replaceable modules, while validation feedback triggers predefined, bounded revisions before the result is saved with a reproducible trace. BS deployment, IRS deployment, and UAV trajectory planning serve as three representative instantiations of this extensible workflow pattern.

Once a validated scene context is available, the normalized deployment workflow is represented as
\vspace{-0.1cm}\begin{equation}
	\mathcal{W}=\left(\mathcal{S}, q, \{(M_j,\theta_j)\}_{j\in\{\mathrm{tar},\mathrm{cand},\mathrm{ckm},\mathrm{opt}\}}\right),
	\label{eq:workflow}
\vspace{-0.1cm}\end{equation}
where $\mathcal{S}$ is the scene context, $q$ is the deployment requirement, $M_j$ is the selected module at stage $j$, and $\theta_j$ is its normalized configuration. Thus, $\mathcal{W}$ is an executable workflow recipe rather than an output artifact. Executing it produces the typed artifacts
\vspace{-0.1cm}\begin{equation}
	(\mathcal{T},\mathcal{C},\mathcal{K},\mathcal{R})=\mathrm{Exec}(\mathcal{W}),
	\label{eq:execution}
\vspace{-0.1cm}\end{equation}
through the serial dataflow
\vspace{-0.1cm}\begin{equation}
	\begin{aligned}
	\mathcal{T} &= M_{\mathrm{tar}}(\mathcal{S}; \theta_{\mathrm{tar}}),\\
	\mathcal{C} &= M_{\mathrm{cand}}(\mathcal{S}, \mathcal{T}; \theta_{\mathrm{cand}}),\\
	\mathcal{K} &= M_{\mathrm{ckm}}(\mathcal{S}, \mathcal{T}, \mathcal{C}; \theta_{\mathrm{ckm}}),\\
	\mathcal{R} &= M_{\mathrm{opt}}(\mathcal{S}, \mathcal{T}, \mathcal{C}, \mathcal{K}; \theta_{\mathrm{opt}}).
	\end{aligned}
	\label{eq:dataflow}
\vspace{-0.1cm}\end{equation}
Here, the scene context $\mathcal{S}$ and the artifacts $\mathcal{T}$, $\mathcal{C}$, $\mathcal{K}$, and $\mathcal{R}$ correspond to the typed contracts \texttt{SceneContext}, \texttt{TargetSpec}, \texttt{CandidateSet}, \texttt{CKMArtifact}, and \texttt{OptimizationResult}. Each execution is associated with a \texttt{WorkflowTrace} containing the normalized workflow, warnings, metrics, and refinement events. \texttt{SceneContext} stores the coordinate frame, bounds, structural and material summaries, and task-aware defaults. \texttt{TargetSpec} defines the evaluation region or route and its service criteria. \texttt{CandidateSet} records admissible decision variables, rejection rules, and sampling metadata. \texttt{CKMArtifact} carries task-indexed channel tensors with their axes, units, and masks. \texttt{OptimizationResult} stores selected variables, objective scores, validation metrics, and visualization data.

Before $\mathcal{S}$ is admitted, deterministic scene validation checks the description, object list, labels, materials, scale, up-axis convention, and exported radio scene for consistency and loadability. Low-risk repairs can regenerate \texttt{SceneContext}, whereas review-level issues are retained in \texttt{WorkflowTrace}. Because this check precedes downstream module execution, target masks and candidate rejection rules are derived from the validated geometry, while CKM construction uses the same geometry together with the validated material assignments.

Validation-guided deployment refinement reuses this interface. After iteration $i$ returns $\mathcal{R}^{(i)}$, validation and guarded refinement are
\vspace{-0.1cm}\begin{equation}
	\begin{aligned}
	\mathcal{V}^{(i)} &= \mathrm{Validate}\left(\mathcal{R}^{(i)},q\right),\\
	\mathcal{W}^{(i+1)} &= \mathrm{Refine}\left(\mathcal{W}^{(i)},\mathcal{V}^{(i)};\gamma\right),
	\end{aligned}
	\label{eq:refinement}
\vspace{-0.1cm}\end{equation}
where $\mathcal{V}^{(i)}$ contains the evaluation metrics, satisfaction status, and diagnostic feedback, and $\gamma$ collects the bounds on node count, candidate count, grid size, compute budget, and constraint relaxation. $\mathrm{Refine}$ is a deterministic rule controller. Without changing the planning objective, it translates identified validation failures or improvement opportunities into bounded adjustments to module choices or configurations, including candidate spacing, optimization budget, path smoothing, and grid resolution. Physical-constraint relaxation is disabled by default and applied only when explicitly authorized; otherwise, the result is flagged for manual review. The artifact contracts remain unchanged, so every refinement pass re-executes the typed workflow according to \eqref{eq:execution} and \eqref{eq:dataflow}.

The interface is designed for extension by separating stable typed artifacts from replaceable module implementations. At the catalog level, each module is identified by its workflow stage, supported service types, and configuration schema, while execution modules exchange only the stage-specific artifacts. The orchestrator can therefore check compatibility without depending on backend internals. Within the proposed framework, ray tracers or measured and learned CKMs can serve as CKM backends when their grids, units, and masks satisfy the same contract; candidate generators and optimizers can be replaced in the same way. A new implementation need only be registered for a stage and satisfy its artifact contract, after which compatible modules can be substituted within an existing service or recombined into a new service without modifying the orchestration layer.


\section{Module Implementation}

The proposed framework is implemented as a local web application with a Python backend and browser interface. Its catalog registers each module together with its configuration schema and stage-specific artifact contract. Given a deployment requirement, the LLM selects compatible modules and instantiates their parameters to compose an executable workflow, which the backend normalizes and validates before execution. The backend executes the selected modules and exposes the resulting deployment, performance metrics, and saved artifacts through the browser interface, while subsequent validation reuses the same workflow and artifact contracts.

\subsection{Scene Construction Module}
The scene construction module accepts two input sources. The user may import an OpenStreetMap (OSM) region \cite{openstreetmap}; the importer downloads a bounded extract, converts closed building ways into local metric meshes, uses height or level tags when available, and adds a ground plane. The user may also upload a stereolithography (STL) or Polygon File Format (PLY) triangle mesh originating from computer-aided design (CAD), public three-dimensional (3D) assets, or reconstruction pipelines.

For both inputs, the module renders a preview and extracts a geometry summary. Mesh components are split into inspectable objects, while OSM footprints and height tags provide the initial object list. Based on the geometry summary and user instructions, the LLM API proposes semantic labels and Sionna-compatible material properties, including material type and thickness, for the scene objects; the user can accept or override them before building the scene.

The scene editor supports table edits and natural-language prompts for component deletion and update, as well as cuboid addition. Prompts such as ``mark the three solid blocks as concrete buildings'' or ``add a medium dry ground slab with 5\% margin'' become explicit edits. Scene validation then compares the instruction with the edited objects, checks description/material consistency, flags suspicious scale or up-axis choices, and returns low-risk repair actions such as filling in missing labels or replacing inconsistent materials. The workflow can apply these actions before export, while issues requiring user review remain visible. The module writes normalized PLY meshes and a Sionna-compatible scene description, verifies that Sionna RT can load the scene, and emits a \texttt{SceneContext}.

\subsection{Target and Candidate Modules}
The target module instantiates the normalized target configuration as a \texttt{TargetSpec} describing the region or route over which performance is evaluated. For BS and IRS deployment tasks, built-in targets are rectangular receiver regions and explicit receiver sets with receiver height, service threshold, and excluded-region handling. Geometry-blocked receiver points, such as those inside buildings, can be excluded from service targets. The configured link-quality metric can be received power, signal-to-noise ratio (SNR), or signal-to-interference-plus-noise ratio (SINR). The coverage rate is the fraction of the nonexcluded target receiver points at which this metric meets or exceeds the service threshold. For the UAV planning task, the target is an ordered set of route control points comprising a start point, a goal point, optional waypoints, and a flight altitude. The feasibility rate is the fraction of evaluated route samples that jointly satisfy the obstacle-clearance and radio-quality constraints.

The candidate module constructs the legal decision space and emits \texttt{CandidateSet}. BS and IRS deployment tasks support region-grid, wall-grid, and explicit-point candidates. Region-grid and explicit-point candidates are filtered for building avoidance and clearance, while the wall-grid sampler selects eligible structures and wall sides, samples them at the specified mounting height and spacing, and applies wall-offset and reference-distance constraints. The UAV planning task uses a geometry-feasible mask to exclude structures and their clearance regions and a radio-feasible mask to enforce the configured link-quality threshold; their intersection defines the joint feasible grid. Candidate outputs record legal and rejected samples, source rules, and sampling metadata.

\subsection{CKM Construction Modules}
The CKM modules convert \texttt{SceneContext}, \texttt{TargetSpec}, and \texttt{CandidateSet} into task-specific \texttt{CKMArtifact}s. For BS deployment, the X2X CKM backend constructs a CGM indexed by each BS candidate-receiver pair. The optimizer then derives received power, SNR, or SINR from the CGM using the configured transmit power, noise, and interference assumptions. For IRS deployment, the cascaded model quantizes the element phases of each panel to focus on a selected target cell. Element fields are coherently combined within each panel, while the powers contributed by the direct path and different panels are added noncoherently. The resulting artifact stores effective phase-focused cascaded-gain CKMs rather than a full per-element channel-vector database. For UAV planning, the backend constructs one B2X CGM per serving BS over the flight-altitude grid and derives the SNR or SINR values used to identify radio-feasible cells. Each \texttt{CKMArtifact} includes tensor shapes, axes, units, valid-cell masks, backend metadata, and storage paths, allowing optimizers to consume CKM values independent of their source.

\subsection{Optimization, Validation, Reuse, and Refinement}
The optimizer consumes a \texttt{CKMArtifact} and returns an \texttt{OptimizationResult}; its configuration specifies the planning objective, node budget, solver, and compute budget. For BS and IRS deployment, candidate subsets can be scored by coverage rate or by the mean or minimum of the configured link-quality metric. The exhaustive solver evaluates all legal candidates for a single-node deployment, the greedy solver iteratively adds the candidate with the largest marginal improvement, and the greedy-swap solver tests one-for-one replacements until no improvement is found or the round limit is reached. An automatic selector chooses among them based on node count, candidate-set size, and compute budget. For UAV planning, the joint geometry- and radio-feasible grid is eroded to enforce the requested obstacle clearance, and the start, waypoints, and goal are snapped to feasible cells within a bounded radius. A$^\ast$ then solves consecutive route legs using physical travel distance as the edge cost. Optional shortcut smoothing removes grid-induced bends only when each replacement segment remains feasible.

After optimization, the evaluation determines whether the \texttt{OptimizationResult} satisfies the original deployment requirement. For BS and IRS deployment, the evaluator compares the achieved coverage rate and any specified mean or minimum link-quality value against their respective targets. For UAV planning, the primary target is a route through all required control points that satisfies obstacle-clearance and radio-quality constraints throughout; route length and turn count are secondary quality measures. If a target is missed or a secondary objective can be improved within the refinement budget, predefined rules make bounded changes to the optimization or sampling parameters and rerun the solver. BS and IRS refinement can densify candidate sampling and increase greedy-swap rounds under fixed node budgets, while UAV refinement can enable shortcut smoothing and refine the grid within a cell-count guard. Retries that exceed guards or require constraint relaxation are marked for user review. Each run persists the workflow, artifacts, validation, and provenance metadata for inspection and reuse.

\section{Case Study and Preliminary Evaluation}

We ran the three representative service instantiations on a compact urban-block scene. All LLM-assisted scene-preparation and initial-workflow-composition steps used \texttt{gpt-5.5} \cite{openai_gpt55}. Building footprints were obtained from OSM and converted to a triangle mesh, and the resulting geometry was reused for BS, IRS, and UAV planning. Unless otherwise stated, ray tracing used a carrier frequency of 3.5 GHz, isotropic single-element arrays, a noise floor of -90 dBm, a random seed of 42, and a maximum path depth of 3.
LoS, diffuse reflection, and refraction were enabled while specular reflection and diffraction were disabled. 
CKMs were constructed on grids with 2 m spacing using 200\,000 ray samples per transmitter.
BS and UAV cases used 10 dBm transmit power; the IRS case used a 20 dBm serving BS and IRSs with 10$\times$10 reflecting elements and four phase levels. We compared validation-guided refinement with the corresponding one-shot workflows. Refinement preserved the original task specification and physical constraints while adaptively increasing sampling resolution and optimization effort. The comparison thus characterizes the complete guarded refinement loop under its bounded compute-allocation policy.

\subsection{Scene Construction}

\begin{figure}[!tb]
	\centering
	\begin{minipage}[b]{1.0\columnwidth}
		\centering
		\includegraphics[width=\linewidth]{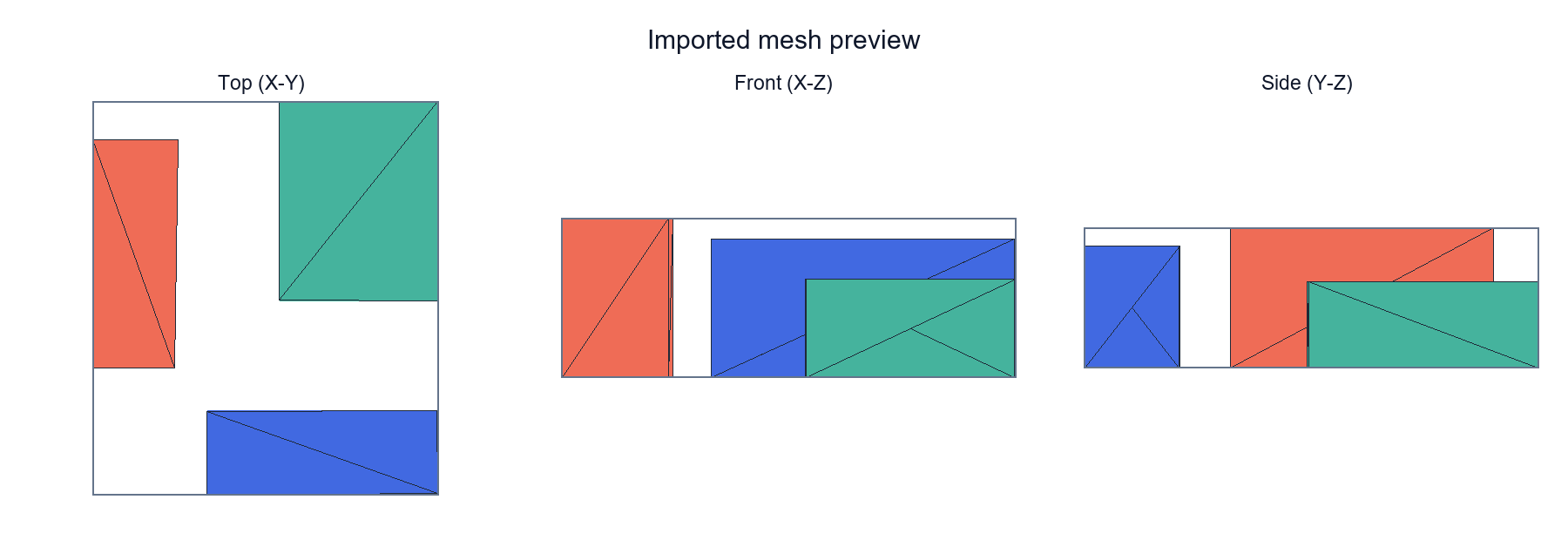}
		\footnotesize\par (a) Imported mesh preview.
	\end{minipage}
	\hfill
	\begin{minipage}[b]{0.39\columnwidth}
		\centering
		\includegraphics[width=\linewidth]{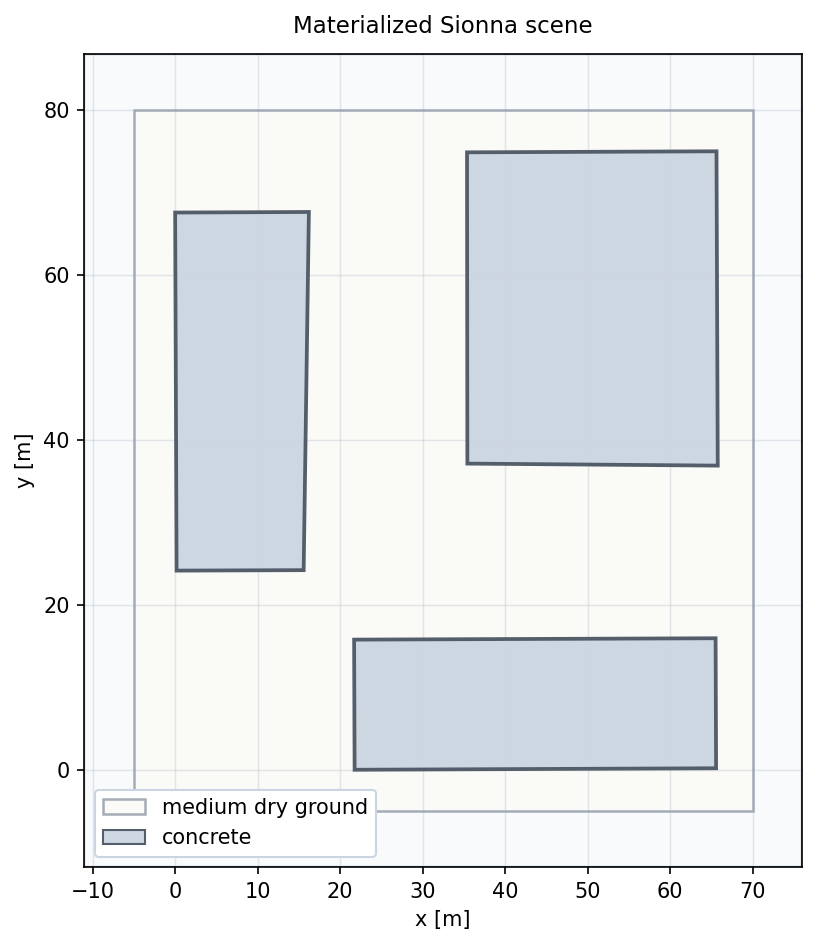}
		\footnotesize\par (b) Materialized Sionna scene.
	\end{minipage}
	\caption{Scene construction output for the urban-block case: imported mesh preview, concrete building solids, and a medium-dry-ground slab.}
	\label{fig:scene_workflow}
\vspace{-0.2cm}\end{figure}

Fig.~\ref{fig:scene_workflow} shows the scene-construction output used by all planning tasks, including the imported mesh preview and the materialized Sionna scene. The prompt was: \emph{Use the outdoor block mesh, assign concrete material to all buildings, and add a ground plane over $x\in$ [-5, 70] m and $y\in$ [-5, 80] m}. Starting from the OSM-derived triangle mesh, the module split the imported geometry into three building solids, assigned concrete material to them, added a medium-dry-ground slab, ran scene validation, and exported four materialized structures. This step is not merely a visualization front end: it fixes the geometry, material table, and coordinate frame before downstream modules run. Thus, blocked-cell masks, legal candidate filters, and material-dependent ray-tracing outputs all refer to the same checked scene representation. Once the scene passes validation, subsequent CKM and optimization modules inherit a consistent geometry-material contract rather than separate task-specific preprocessing assumptions.

\subsection{Deployment and Trajectory Planning}

\begin{figure}[!tb]
	\centering
	\begin{minipage}[b]{0.49\columnwidth}
		\centering
		\includegraphics[width=\linewidth]{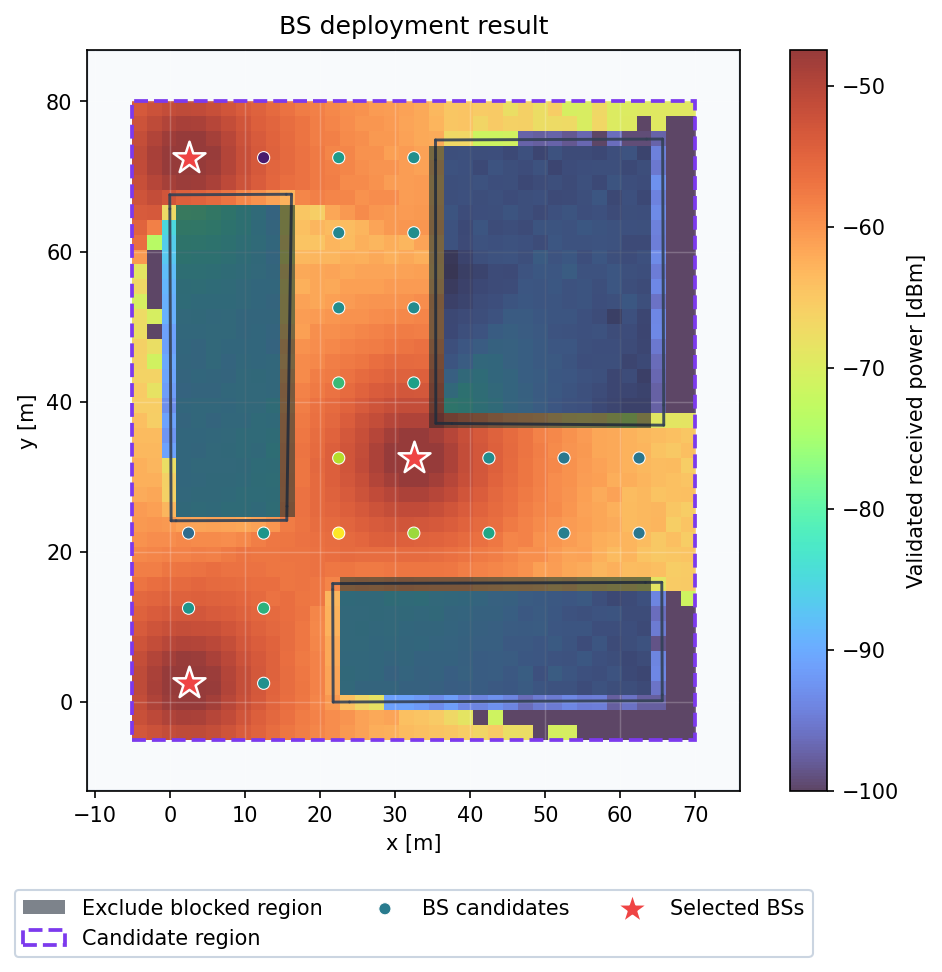}
		\footnotesize\par (a) Refinement off.
	\end{minipage}
	\hfill
	\begin{minipage}[b]{0.49\columnwidth}
		\centering
		\includegraphics[width=\linewidth]{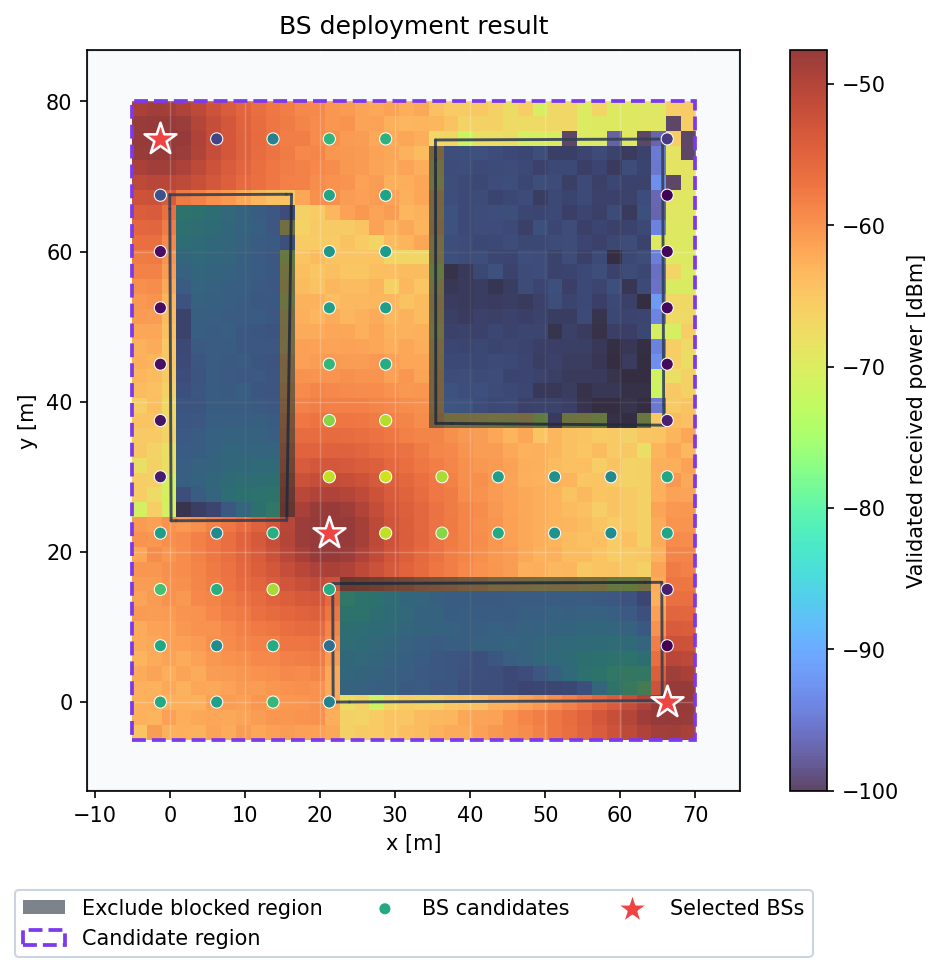}
		\footnotesize\par (b) Refinement on.
	\end{minipage}
	\caption{BS deployment result: the guarded controller densifies candidate sampling from 10 m to 7.5 m, increasing coverage from 84.9\% to 97.8\% with additional search effort.}
	\label{fig:bs_workflow}
\vspace{-0.2cm}\end{figure}

Fig.~\ref{fig:bs_workflow} shows the BS deployment result over the receiver grid. The prompt was: \emph{Place at most three BSs at 5 m height over the whole scene, use 10 m spacing for legal candidate sites, evaluate receivers at 1 m height on a 2 m grid, and reach 90\% coverage above -85 dBm}. With refinement disabled, the workflow instantiated 26 legal BS sites, an X2X \texttt{CKMArtifact} of size 26$\times$43$\times$38, and greedy-swap optimization for three BSs, covering 860/1013 valid target cells. Since this 84.9\% coverage missed the target, two refinement iterations reduced the candidate spacing to 7.5 m and increased the swap budget, producing 59 candidates and a 59$\times$43$\times$38 CKM. The selected BSs were located at (66.25, 0.00, 5.0) m, (21.25, 22.50, 5.0) m, and (-1.25, 75.00, 5.0) m. Final validation covered 991/1013 cells, giving 97.8\% coverage and -61.0 dBm mean received power. The gain came from more legal BS sites and more local-search effort under the same receiver grid and -85 dBm threshold.

\begin{figure}[!tb]
	\centering
	\begin{minipage}[b]{0.49\columnwidth}
		\centering
		\includegraphics[width=\linewidth]{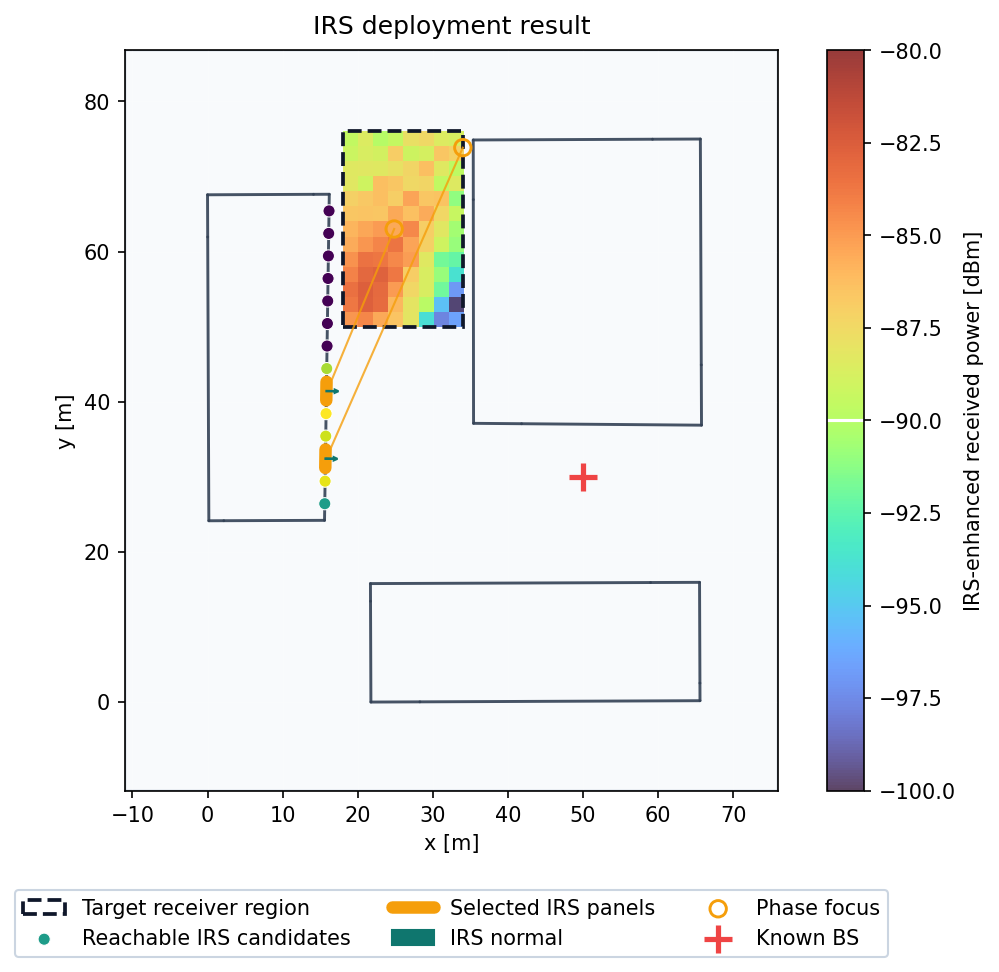}
		\footnotesize\par (a) Refinement off.
	\end{minipage}
	\hfill
	\begin{minipage}[b]{0.49\columnwidth}
		\centering
		\includegraphics[width=\linewidth]{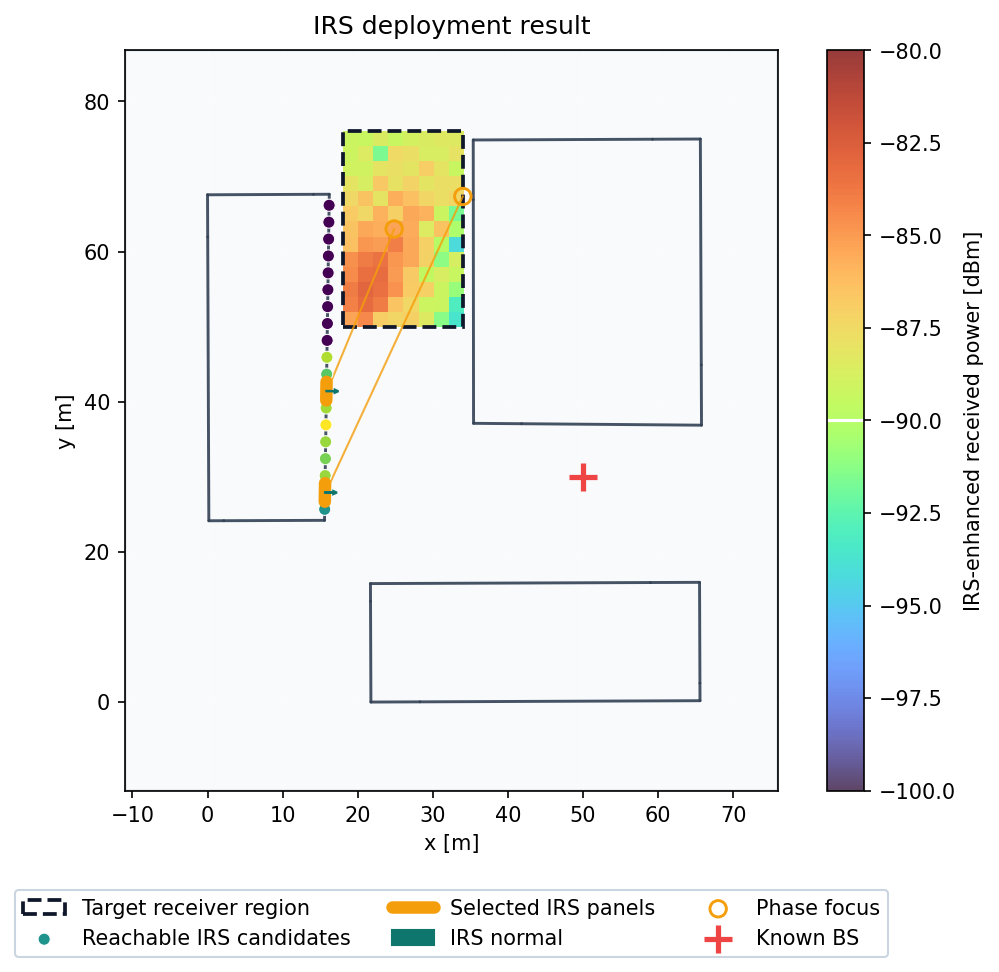}
		\footnotesize\par (b) Refinement on.
	\end{minipage}
	\caption{IRS deployment result: the guarded controller densifies wall-mounted candidates from 3 m to 2.25 m, increasing coverage from 86.5\% to 91.3\% with additional search effort.}
	\label{fig:irs_workflow}
\vspace{-0.2cm}\end{figure}

Fig.~\ref{fig:irs_workflow} shows the candidate wall, selected panels, serving BS, and phase-focusing points for the IRS task. The prompt was: \emph{A BS is fixed at (50, 30, 5) m. Improve the weak region over $x\in$ [18, 34] m and $y\in$ [50, 76] m by mounting at most two IRS panels at 3 m height on the right wall of the left-side building, use 3 m wall spacing, and reach 90\% coverage above -90 dBm}. The initial workflow formed a rectangular \texttt{TargetSpec}, sampled 14 legal wall-mounted candidates, built a 14$\times$13$\times$8 cascaded \texttt{CKMArtifact}, and covered 90/104 cells. Since 86.5\% coverage was below target, two refinement iterations reduced the wall spacing to 2.25 m and increased the swap budget. The revised workflow sampled 19 candidates and selected mounts at (15.64, 27.92, 3.0) m and (15.83, 41.41, 3.0) m, improving coverage to 95/104 cells with -87.4 dBm mean and -94.2 dBm minimum received power. Throughout refinement, the wall and two-panel constraints remained fixed while the wall-surface sampling and local-search resolution were increased.

\begin{figure}[!tb]
	\centering
	\begin{minipage}[b]{0.49\columnwidth}
		\centering
		\includegraphics[width=\linewidth]{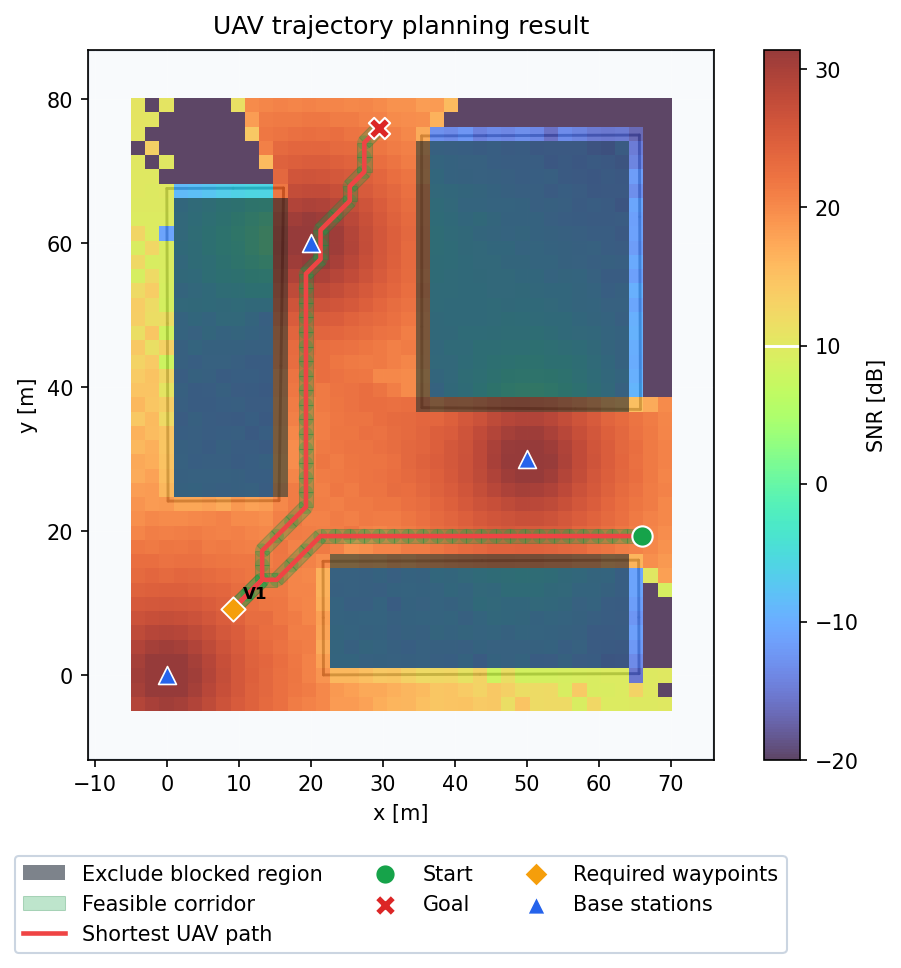}
		\footnotesize\par (a) Refinement off.
	\end{minipage}
	\hfill
	\begin{minipage}[b]{0.49\columnwidth}
		\centering
		\includegraphics[width=\linewidth]{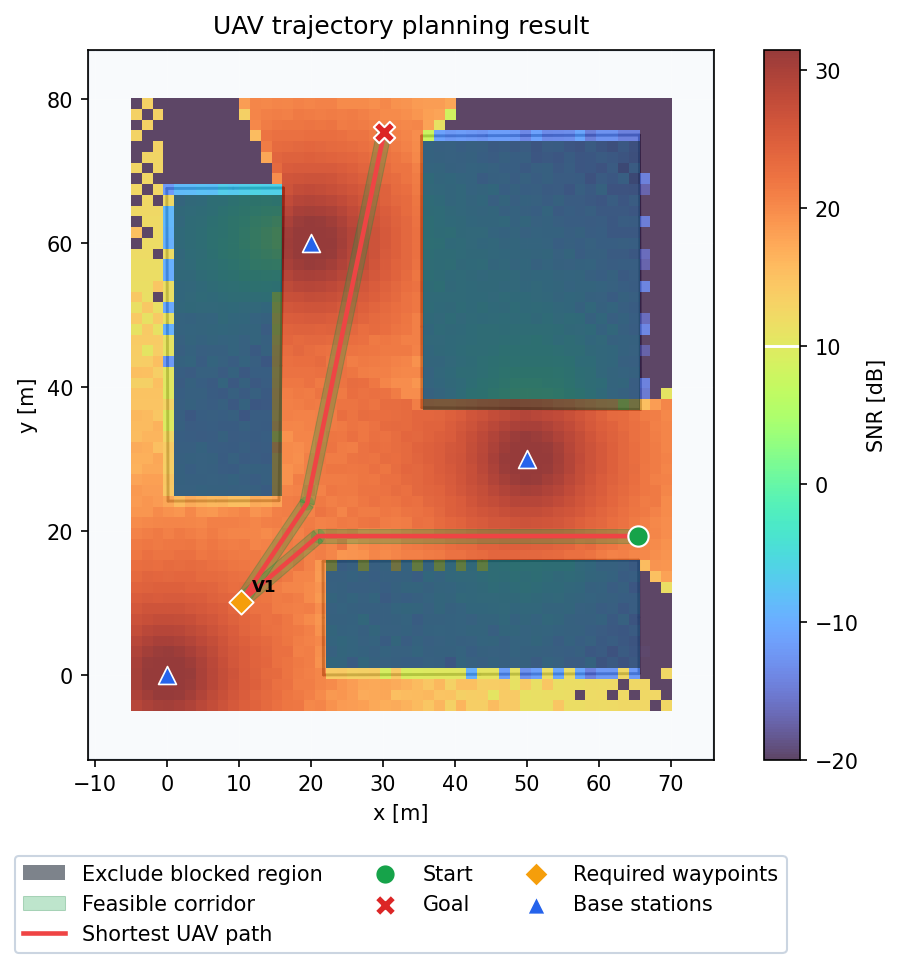}
		\footnotesize\par (b) Refinement on.
	\end{minipage}
	\caption{UAV trajectory planning result: the guarded controller enables shortcut smoothing and 1.5 m grid refinement, shortening the route from 136.1 m to 127.6 m with additional computation.}
	\label{fig:uav_workflow}
\vspace{-0.2cm}\end{figure}

Fig.~\ref{fig:uav_workflow} shows the planned UAV corridor together with the SNR values derived from the CKM. The prompt was: \emph{Plan a constant-altitude route at 10 m from (65, 20) m to (30, 75) m through the required waypoint (10, 10) m. Use BSs at (50, 30, 5) m, (0, 0, 5) m, and (20, 60, 5) m, maintain at least 2 m clearance from obstacles, require SNR above 10 dB, and smooth the route}. With refinement disabled, the workflow intersected the geometry-feasible and SNR-feasible masks, eroded the result for the 2 m corridor, built a 3$\times$43$\times$38 multi-BS \texttt{CKMArtifact}, and found a 136.1 m A$^\ast$ route with 62 path vertices and 14 turns. Across three refinement iterations, the rule controller enabled shortcut smoothing and refined the grid from 2 m to 1.5 m, producing a 3$\times$57$\times$50 CKM, 1119 centerline-feasible cells, and a 127.6 m route with 5 path vertices and 3 turns. The minimum SNR remained 20.1 dB, above the 10 dB threshold. Thus, geometric and CKM-based radio feasibility were fused in one graph search, and the shorter route was obtained without relaxing the original SNR or obstacle-clearance constraints.

\subsection{Preliminary Comparison}

\begin{table}[!tb]
	\caption{Preliminary Comparison on the Urban-Block Scene}
	\label{tab:prelim}
	\centering
	\scriptsize
	\renewcommand{\arraystretch}{0.92}
	\setlength{\tabcolsep}{1.6pt}
	\begin{tabular}{|p{0.06\linewidth}|p{0.35\linewidth}|p{0.17\linewidth}|p{0.35\linewidth}|}
		\hline
		\textbf{Task} & \textbf{Method} & \textbf{Cov./feas.} & \textbf{Quality measures} \\
		\hline
		BS & Random three-BS & 73.1\% covered & Mean/min -74.3/$<$-110 dBm \\
		\hline
		BS & Greedy three-BS & 84.8\% covered & Mean/min -67.2/$<$-110 dBm \\
		\hline
		BS & Greedy-swap, refinement off & 84.9\% covered & Mean/min -65.9/$<$-110 dBm \\
		\hline
		BS & Greedy-swap, refinement on & 97.8\% covered & Mean/min -61.0/$<$-110 dBm \\
		\hline
		IRS & Direct-only baseline & 0\% covered & Mean/min $<$-110/$<$-110 dBm \\
		\hline
		IRS & Random single panel & 49.0\% covered & Mean/min -90.6/-108.7 dBm \\
		\hline
		IRS & Best single panel & 51.0\% covered & Mean/min -91.4/$<$-110 dBm \\
		\hline
		IRS & Greedy-swap, refinement off & 86.5\% covered & Mean/min -87.6/-101.4 dBm \\
		\hline
		IRS & Greedy-swap, refinement on & 91.3\% covered & Mean/min -87.4/-94.2 dBm \\
		\hline
		UAV & Straight polyline & 56.5\% feasible & 127.4 m; min SNR -9.2 dB \\
		\hline
		UAV & Radio-only A* & 73.3\% feasible & 136.1 m; min SNR 15.5 dB \\
		\hline
		UAV & CKM-aware A*, refinement off & 100\% feasible & 136.1 m; min SNR 20.1 dB \\
		\hline
		UAV & CKM-aware A*, refinement on & 100\% feasible & 127.6 m; min SNR 20.1 dB \\
		\hline
	\end{tabular}
	\par Note: cov. = coverage rate; feas. = feasibility rate.
\vspace{-0.2cm}\end{table}

Table~\ref{tab:prelim} reports a preliminary comparison on the same scene. Baselines reused the same \texttt{TargetSpec}, \texttt{CandidateSet}, and \texttt{CKMArtifact} as the corresponding one-shot workflow whenever applicable; refinement regenerated candidate and CKM artifacts only after bounded revisions. For BS deployment, random, greedy, and one-shot greedy-swap represent unguided selection, basic CKM-aware selection, and local search, respectively, while refinement resolves the limiting candidate density. For IRS deployment, the direct-only and single-panel results motivate the cascaded two-panel formulation, and refinement increases the resolution of legal wall sampling and local search. For UAV planning, the straight polyline ignores radio and collision constraints, radio-only A$^\ast$ satisfies the SNR threshold but can intersect buildings, and CKM-aware A$^\ast$ satisfies both. 
With the additional bounded sampling and search effort specified above, the guarded loop addresses the distinct bottlenecks in BS candidate density, IRS wall sampling, and UAV route discretization without changing the service or feasibility constraints. These results establish the end-to-end operation of the proposed workflow on the representative urban scene; evaluation across larger scenes and task distributions remains future work.

\balance
\section{Conclusion}
This paper has presented an LLM-orchestrated agent framework for CKM-assisted network deployment optimization that organizes deployment services through a typed, modular, and reproducible workflow paradigm. The LLM layer supports intent interpretation and initial workflow composition; physics-grounded modules produce numerical artifacts; and a deterministic guarded controller performs bounded validation-guided refinement. The urban-block case study demonstrates reuse of the common interface for BS, IRS, and UAV planning services. Future work will enable the LLM agent to generate and validate new planning modules from user requirements, extending the framework beyond the currently instantiated services.

\bibliographystyle{IEEEtran}
\bibliography{myref}

\end{document}